\documentclass[12p, trans]{IEEEtran}

\usepackage{subfigure}
\usepackage[dvipsnames]{xcolor}
\usepackage{amsmath}
\usepackage{amssymb}
\usepackage{amsthm}
\usepackage{comment}
\usepackage{color}
\usepackage{mathtools}
\usepackage{algorithm}
\usepackage{algorithmic}
\usepackage{datetime}
\usepackage{graphicx}
\usepackage{cuted}
\usepackage{cite}

\ifCLASSINFOpdf
\else
\fi
\begin{document}
%

\title{5G New Radio Unlicensed: Challenges and Evaluation}

\author{\IEEEauthorblockN{Mohammed Hirzallah$^1$, Marwan Krunz$^{1}$, Balkan Kecicioglu$^2$ and Belal Hamzeh$^2$}

\IEEEauthorblockA{$^1$Department of Electrical and Computer Engineering, University of Arizona, AZ, USA\\
$^2$CableLabs, Louisville, CO, USA\\
Email: \{hirzallah, krunz\}@email.arizona.edu}, \{b.kecicioglu, b.hamzeh\}\ @cablelabs.com 
}


%


\maketitle


\begin{abstract}
To meet the high demand for mobile data, the Third Generation Partnership Project (3GPP) established a set of standards known as 5G New Radio (5G NR). The architecture of 5G NR includes a flexible radio access network and a core network. 3GPP has also been working on a new radio access technology, called 5G NR Unlicensed (5G NR-U), which aims at extending 5G NR to unlicensed bands. In this paper, we give an overview of the most recent 5G NR-U design elements and discuss potential concerns, including fair coexistence with other unlicensed technologies such as Wi-Fi. We use simulations to study coexistence between Wi-Fi and 5G NR-U systems. Our evaluation indicates that NR-U often achieves higher throughput and lower delay than Wi-Fi (802.11ac). The two systems experience different buffer occupancies and spectrum utilization statistics. We also discuss the improvements that NR-U offers over LTE Licensed Assisted Access (LTE-LAA). 
\end{abstract}


\section{Introduction}
Next-generation wireless networks will support applications with widely diverse performance requirements. In its International Mobile Communications (IMT)-2020 recommendations,  the International Telecommunications Union (ITU) specifies three use cases for next-generation wireless networks: Enhanced mobile broadband (eMBB), ultra-reliable and low latency communication (URLLC), and massive machine-type communication (mMTC). While these use cases embody different performance requirements, they all share the need for more spectrum. In its effort to extend 5G cellular operation to unlicensed spectrum, 3GPP is initially targeting  the Unlicensed National Information Infrastructure (UNII) bands at 5 GHz and 6 GHz. Future specifications will address unlicensed millimeter wave (mmWave) bands at 60 GHz. Wireless systems can operate over unlicensed bands as long as they comply with spectrum regulations, which are intended to ensure harmonious coexistence of various incumbents that operate on the same band. The ubiquity of Wi-Fi networks makes achieving harmonious 5G NR-U and Wi-Fi coexistence a key objective for NR-U designers. To ensure fairness in channel access, NR-U should not impact an existing Wi-Fi system more than the impact of another Wi-Fi system \cite{NR-U-WID}. 

Early works surveying 5G NR-U can be found in  \cite{Semaan2017wcnc-nru-beamforming,Pater2018mag-5g-laa-slicing-huwaei,Lagen2020nru-mmwave,Oh2019globecom-nru-samsung}. These works focused on pre-standard NR-U operation at sub-6 GHz and/or mmWave frequencies and discussed the feasibility of utilizing the channel access procedures of `further enhanced' LTE LAA (feLAA) in 5G networks. The effectiveness of unlicensed bands for IoT applications was investigated in \cite{Lu2019nru-Iot}, where the authors studied challenges associated with extending 5G services to unlicensed bands. Recently, 3GPP added more details and features to the NR-U specifications, including new deployment scenarios as well as other enhancements, such as interlace waveform design, multi-channel operation, frequency reuse, and initial access \cite{TR38.889-nru}. Another work that focused on studying Physical-layer aspects of NR-U can be found in \cite{zhao2019nru-phy}. The authors in \cite{Oh2019globecom-nru-samsung} investigated the adaptation of the contention window for NR-U. \textcolor{black}{Analysis and evaluation of latency and reliability in 5G NR-U were discussed in \cite{Maldona2020-access}, where the authors suggested modifications to improve both metrics. Evaluation of different aspects of coexistence between NR-U and IEEE 802.11ad-based Wi-Fi at mmWave frequencies, including fairness and setting of detection thresholds, was provided in 
\cite{Patriciello2020-access}.} \textcolor{black}{Machine learning techniques to mitigate interference between NR-U operators and improve spatial reuse in NR-U/Wi-Fi coexistence were presented in \cite{hirzallah2020-dissertaion}\cite{Hirzallah2019Secon-matchmaker}. Authors in \cite{kosek2020downlink} analyzed NR-U/Wi-Fi coexistence analytically and concluded that novel mechanisms are still needed to improve the fairness over unlicensed bands.}

In this paper, we provide an overview of the most recent NR-U specifications and discuss various deployment options. We present one possible radio stack architecture for embedding 5G NR-U capabilities in future NR designs. We also investigate the challenges associated with NR-U PHY, MAC, and upper layers so as to achieve harmonious NR-U/Wi-Fi coexistence over the unlicensed 5 GHz and 6 GHz bands. Simulation-based evaluation of User Perceived Throughput (UPT), latency, buffer occupancy, and spectrum utilization are provided for indoor and outdoor scenarios in both bands. The rest of the paper is organized as follows. In Section \ref{sc:overview}, we provide an overview of cross-technology coexistence over unlicensed spectrum. In Section \ref{sc:nru-overview}, we introduce the NR-U design. In Section \ref{sc:nru-challenges}, we discuss key challenges affecting the harmonious coexistence between NR-U and Wi-Fi networks. We present our evaluation for NR-U/Wi-Fi coexistence in Section \ref{sc:evaluation} and conclude in Section \ref{sc:conclusions}.

\section{Coexistence of Heterogeneous Technologies Over Unlicensed Bands} \label{sc:overview}

\subsection{5G NR-U Frequency Bands} \label{sc:spectrum-bands}

\begin{figure} 
\centering
\includegraphics[scale=0.575]{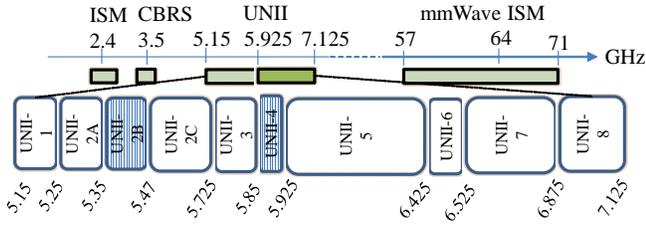}
\caption{Unlicensed/shared spectrum bands for NR-U operation (unlicensed operation over UNII-2B and UNII-4 bands is restricted).} \label{fig:spectrum}
\end{figure}

As shown in Figure \ref{fig:spectrum}, two frequency ranges are targeted for NR-U operation: Low-frequency bands below $7$ GHz and a high-frequency band at $60$ GHz. Specifically, about $2$ GHz of unlicensed/shared spectrum is available for omni-directional communications below $7$ GHz over the Industrial Scientific Medical (ISM) band at $2.4$ GHz, the Citizens Broadband Radio Service (CBRS) band at $3.5$ GHz, and the UNII bands at $5$ GHz and $6$ GHz frequencies \cite{FCC-6GHz-18-295}. There is also $14$ GHz of unlicensed spectrum available at the $60$ GHz band that can be used for directional communications \cite{FCC-60GHz-340310A1}.

\textcolor{black}{FCC has just recently announced its proposed rule making to open up bands from $5.925$ GHz to $7.125$ GHz for unlicensed access under part 15 rules \cite{FCC-6GHz-18-295}. Different UNII bands have different restrictions on the maximum transmit power, effective isotropic radiated power (EIRP), applicability for indoor/outdoor operation, and the requirement for dynamic frequency selection (DFS). Unlicensed users are required to perform DFS to avoid interference with radars and other licensed services operating in UNII bands. Under DFS, an unlicensed device has to interrupt its transmission and perform periodic sensing of radar signals.  \textcolor{black}{When a radar signal is detected, transmission should be stopped within 10 seconds and channel should be abandoned for 30 minutes. Detection methods of radar signals are not specified and left for implementation.}  The $5$ GHz  band is divided into non-overlapping channels of $20$ MHz bandwidth. Wider channels (e.g., 40, 80, and 160 MHz) can be constructed via channel bonding. NR-U systems are allowed to coexist with IEEE 802.11n/ac/ax-based as well as LTE-LAA services over these channels.}

As for the $6$ GHz band, much of it is currently occupied by some licensed services, including point-to-point microwave links, fixed satellite systems, and mobile services, such as the broadcast auxiliary service and the cable TV relay service. To protect these licensed services, unlicensed users are also required also to perform \emph{automatic frequency coordination (AFC)}, where protection zones are established around the incumbent services and unlicensed users are not allowed to access bands in these protection zones. Unlicensed users are also required to control their transmit power and restrict their transmission to indoor whenever AFC fails \cite{FCC-6GHz-18-295}.  In the $6$ GHz band, NR-U is expected to coexist with IEEE 802.11ax/be-based systems (Wi-Fi 6/Wi-Fi 7). \textcolor{black}{The new FCC rules allow unlicensed operation over most of the UNII bands in the 5.925 - 7.125 GHz range. To protect incumbent services, the FCC restricts the transmit power of outdoor base stations to 23 dBm/MHz in addition to still performing AFC. The EIRP over 320 MHz channel bandwidth (the maximum channel bandwidth) should not exceed 36 dBm. Outdoor user equipments (UEs) can transmit up to 17 dBm/MHz but subject to 30 dBm EIRP limit on 320 MHz channel bandwidth. Outdoor operation is limited to UNII-5 (5.925–6.425 GHz) and UNII-7 (6.525–6.875 GHz) bands. Indoor operation can take place over all UNII bands, i.e., UNII-5/-6/-7/-8, and AFC is not required. However, indoor base stations are limited to 5 dBm/MHz and they are subject to maximum of 30 dBm EIRP limit on 320 MHz channel bandwidth. Indoor UEs can transmit at $-1$ dBm/MHz without exceeding the 24 dBm EIRP limit over 320 MHz channel bandwidth \cite{FCC_2020-6ghz}}. \textcolor{black}{3GPP has recently kicked off the study of licensed and unlicensed NR operation over 6 GHz bands \cite{TR37.890-6ghz}. Authors in \cite{naik2020next} discussed some challenges associated with wireless operation over unlicensed 6 GHz bands.}

\begin{figure*}
\centering
\includegraphics[scale=0.48]{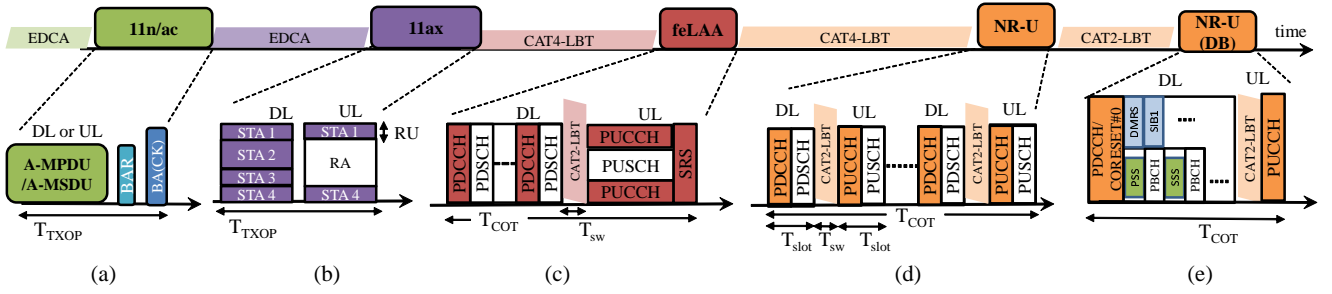}
\caption{Operation of different technologies over unlicensed UNII bands below 7 GHz, (a) IEEE 802.11n/ac, (b) IEEE 802.11ax, (c) LTE-LAA exemplified by feLAA, (d) NR-U without discovery transmission, and (e) NR-U with discovery frame transmission.} \label{fig:arbitraryTimeOperation}
\end{figure*}

\subsection{Operation of Incumbent Systems}
NR-U-based systems will primarily share the unlicensed UNII bands below $7$ GHz with LTE-LAA-based and with IEEE 802.11-based systems. To operate over the UNII bands, these systems rely on different channel access procedures, all of which require sensing the channel before transmission. This mechanism is called Listen-Before-Talk (LBT), a flavor of Carrier Sense Multiple Access with Collision Avoidance (CSMA/CA). In CSMA/CA with \emph{exponential backoff}, a device backs off for $k$ idle slots. A channel is deemed to be idle if it remains so for an Arbitration Inter-frame Space (AIFS) duration ($T_\text{AIFS}$), a.k.a., \emph{defer time} ($T_\text{df}$). To reduce the possibility of a collision, devices need to back off for different $k$ values. Accordingly, $k$ is sampled randomly from the range $ \{0,\cdots, W_{j}-1\}$, where $W_{j} = \min \{2^j \text{CW}_{\min}, \text{CW}_{\max}\}$. $\text{CW}_{\min}$ is the minimum contention window, $\text{CW}_{\max}$ is the maximum contention window, and $j$ is the index of the retransmission attempt. If the transmission fails, the device doubles its contention window size. The values of CW\textsubscript{min}, CW\textsubscript{max}, and AIFS impact the channel access delay and collision rate of coexisting devices.  After contending for $k$ idle slots, a device can use the channel for a time period known as \emph{channel occupancy time (COT)}, which is referred to as \emph{transmit opportunity (TXOP)} period in IEEE 802.11-based systems. Coexisting technologies differ in their CSMA/CA parameters as well as on how they leverage their airtime. They also differ in their reaction to failed/collided transmissions, as discussed next.

\begin{table}[tbp]
  \centering
  \caption{EDCA channel access parameters for different ACs \cite{IEEE802.11-2016}}
  \label{tb:AC}
  
  \begin{subtable}
  {}
    \begin{tabular}{|l|l|l|l|l|}
  \hline
  \hline
  AC $A_i$ & $d_i$/$T_{\text{AIFS}}$& $\text{CW}_{\min}$ & $\text{CW}_{\max}$ & Max TXOP $T_{i}$ \\
  \hline
  \hline
  AC\_VO &2/ $34\:\mu$sec & 4 & 8  & 2.08 msec \\
  AC\_VI & 2/ $34\:\mu$sec & 8 & 16  & 4.096 msec \\
 AC\_BE & 3/ $43\:\mu$sec & 16 & 1024  & $-^*$ \\
  AC\_BK & 7/ $79\:\mu$sec & 16 & 1024  & $-$ \\
  Legacy DCF & 2/ $34\:\mu$sec& 16 & 1024  & $-$ \\

  \hline
  \end{tabular}  
  
  \end{subtable}
\begin{tabular}{|l|}
$*$For fair comparison, we set TXOP for AC\_BE to 8 milliseconds\\ in our simulations.\\
  \hline
  \hline
\end{tabular}  

  \end{table}

\begin{table}[tbp]
  \centering
  \caption{CAT4-LBT channel access parameters for different PCs \cite{TR36.213LAA}\cite{TS37.213-nru}}
\label{tb:PC}
  \begin{subtable}{}
  \begin{tabular}{|l|l|l|l|l|}
  \hline
  \hline
  PC $P_i$ &$d_i$/$T_{\text{df}}$&$\text{CW}_{\min}$ & $\text{CW}_{\max}$& Max. COT $T_{i}$ \\
  \hline
  \hline
  $P_1$ & 1, 2/ $25$, $34$ $\mu$sec & $4$ & $8$ & $2$ msec \\
  $P_2$ & 1, 2/ $25$, $34$ $\mu$sec & $8$ & $16$ & $3$ or $4$ msec  \\
  $P_3$ & 3/ $43$ $\mu$sec & $16$ & $64$ & $6$, $8$, $10$ msec   \\
  $P_4$ & 7/ $79$ $\mu$sec & $16$ & $1024$ & $6$, $8$, $10$ msec  \\
  \hline
  \end{tabular}
\end{subtable}

  \end{table}

\subsubsection{IEEE 802.11-based Systems}

IEEE 802.11-based systems (i.e., Wi-Fi) use the Enhanced Distributed Channel Access (EDCA) scheme to coordinate channel access among Wi-Fi devices. EDCA is based on CSMA/CA with exponential backoff. It supports four \emph{access categories (ACs)} for voice, video, best effort, and background traffic. Each AC is associated with a set of contention parameters, shown in Table \ref{tb:AC}. During a TXOP, multiple MAC Service/Packet Data Units can be aggregated and acknowledged via a single \emph{Block ACK} (BA) frame, as shown in Figure \ref{fig:arbitraryTimeOperation}(a). \textcolor{black}{ The transmitter sends a \emph{block ACK request} (BAR) frame, which triggers the receiver to reply back with a BA frame. Two BA policies can be configured: \emph{Immediate BA} and \emph{delayed BA}. Under the immediate BA policy, the receiver should send a BA frame right after the end of the TXOP, while in the delayed BA policy, the receiver can postpone sending the BA and can acknowledge multiple TXOPs using a single delayed BA frame. The delayed BA policy is added to support delay-tolerant applications and reduce their control overhead. Under this policy, the transmitter should have enough buffering capabilities to account for outstanding frames that are not acknowledged.} The IEEE 802.11 standards also support channel bonding and aggregation, as well as single-user MIMO and downlink multi-user MIMO (MU-MIMO) communications.

 The IEEE 802.11ax amendment adds more features to improve frequency reuse and support higher network efficiency. For example, the TXOP can be split between uplink (UL) and downlink (DL), as shown in Figure \ref{fig:arbitraryTimeOperation}(b). Similarly, in the frequency domain, the channel can be divided into several resource units (RUs), enabling Orthogonal Frequency Division Multiple Access (OFDMA). An RU is basically a set of contiguous subcarriers. It is also possible to poll STAs to start UL transmission by sending them a special trigger frame. \textcolor{black}{Some uplink RUs can be dedicated for allowing random access (RA) by stations, a.k.a., \emph{OFDMA BackOff (OBO) procedure}.} \textcolor{black}{Additional details of IEEE 802.11ax operation can be found in \cite{Khorov2019tutorial-11ax}.} A study group of the extremely high throughput Wi-Fi, a.k.a., IEEE 802.11be and Wi-Fi7, has been working on further boosting the performance offered by IEEE 802.11ax systems by incorporating new features and enhancements, including full-duplex communications \cite{Hirzallah2018-tccn-fd-wifi}, multi-channel/multi-band operation, support for wider channel and MIMO communications with larger number of antennas, support of higher modulation schemes, coordination between access points, multi-RU operation, enhancements to link adaptation and retransmission, preamble puncturing, etc \cite{Lopez2019comm-mag-11be, Deng2020-11be-tutorial, Khorov2020-access-11be}.

\subsubsection{LTE-LAA-/NR-U-based Systems} \label{sc:lte-felaa-nr-u-overview}
To facilitate 5G NR-U (also LTE-LAA) operation over unlicensed bands, four LBT \emph{Categories (CATs)} have been defined:
\begin{itemize}
\item CAT1-LBT \textcolor{black}{(Type 2C)}: A gNB can access the channel immediately without performing LBT. The COT can be up to $584$ microseconds.  
\item CAT2-LBT \textcolor{black}{(Type 2A and 2B)}: An NR-U device must sense the channel for a fixed time duration, $T_{\text{fixed}}$. If the channel remains idle during this period, the device can access the channel. \textcolor{black}{In Type 2A, $T_{\text{fixed}}$ is $25$ microseconds, while in Type 2B, it is $16$ microseconds.}
\item CAT3-LBT: An NR-U device must back off for a random period of time before accessing the channel. This random period is sampled from a fixed-size contention window. \textcolor{black}{The option of CAT3-LBT has been excluded from the specifications.}
\item CAT4-LBT \textcolor{black}{(Type 1)}: An NR-U device must back off according to the CSMA/CA procedure with exponential backoff.
\end{itemize}

CAT4-LBT is already adopted by LTE-LAA and is also considered as the baseline NR-U operation for shared spectrum access or \emph{Load Based Equipment} (LBE). \textcolor{black}{Contention window adjustment of LAA has been adopted as the baseline for NR-U. Feedbacks of acknowledgement for a reference subframe (usually the first subframe in a COT) are monitored to decide on doubling the contention window size. If the number of NACK exceeds a threshold (usually 80\%), the contention window size is doubled (more details can be found in \cite{TS37.213-nru}).} Multiple \emph{priority classes} (PCs) are available for different traffic types, similar to EDCA (see Table \ref{tb:PC}). \textcolor{black}{In 3GPP PCs are also referred to as channel access priority classes (CAPCs). Defer time, $T_{\text{df}}$, of DL is set smaller than UL to give DL higher priority to access channel than UL}. The interplay between traffic classes and their effective throughput and average contention delay have been investigated in \cite{Hirzallah2019tccn-modeling}\cite{Hirzallah2018Dyspan-laa-modeling}. Authors in \cite{sathya2020measurement} also conducted real measurements to evaluate the performance of LAA PCs under different traffic profiles in Chicago area. During a COT, multiple DL and UL occasions can be initiated in which UEs are assigned to different resources that are distributed in time, frequency, and spatial domains. 

 \textcolor{black}{CAT2-LBT is used for semi-static channel access, a.k.a., \emph{Frame Based Equipment} (FBE), or to send critical frames, such as discovery frames, or to access channel when it is not shared by others. In these cases, $T_{\text{fixed}}$ can be as small as 9 microseconds. FBE operation mandates a duty cycle of 1/20 and the channel should not be accessed for a while after the end of COT (at least 5 percent of the COT duration).} CAT2-LBT is also required if the time to switch between DL and UL exceeds a certain limit, i.e., $16$ microseconds. LTE-LAA and NR-U differ in their timing resolution, number of possible UL and DL occasions during a COT, as well as their Hybrid Automatic Repeat reQuest (HARQ) designs. \textcolor{black}{In LTE-LAA, the eNB (the designation of base station in LTE) initiates a COT by contending according to CAT4-LBT, as shown in Figure \ref{fig:arbitraryTimeOperation}(c). feLAA enhances the baseline LAA design by adding additional features, such as support of switching between DL and UL within the same COT and support of autonomous uplink. \textcolor{black}{In NR-U, its is possible to have multiple switching occasions between DL and UL (and vice versa) for gNB-initiated COT. For UE-initiated COT, switching is allowed only from UL to DL, and the DL is used to send control signaling.} Once the eNB/gNB reserves the channel, it sends a downlink frame that consists of a sequence of Physical Downlink Control Channels (PDCCHs) and Physical Downlink Shared Channels (PDSCHs). The PDCCH includes the control information needed by UEs to decode their data messages in the PDSCH. In the uplink part of the COT, UEs can send their control and data messages as part of the Physical Uplink Control Channel (PUCCH) and Physical Uplink Shared Channel (PUSCH), respectively. UEs can also send a sounding reference symbol (SRS), which can be used for uplink channel quality estimation for a wider bandwidth.}

\section{NR-U Design Principles} \label{sc:nru-overview}
\subsection{Deployment Options} \label{sc:deployment-options}

\begin{figure*}
\centering
\includegraphics[scale=0.48]{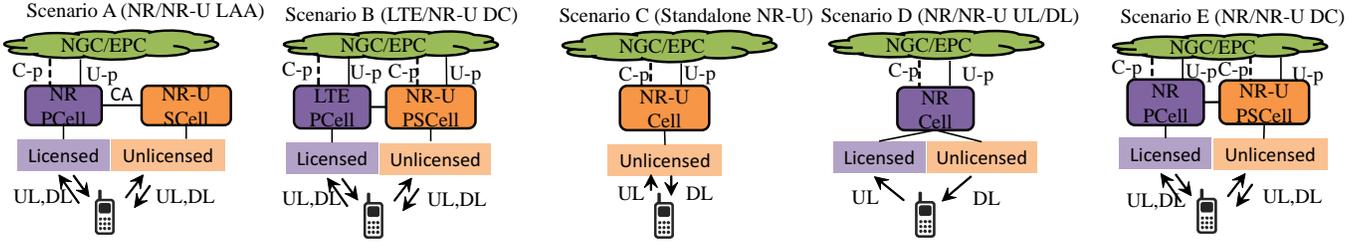}
\caption{NR-U deployment scenarios (`C-p': Control plane; `U-p': User plane; NGC: Next-Generation Core; EPC: Evolved packet core).} \label{fig:deploymentOptions}
\end{figure*}

 \emph{Dual connectivity} (DC) and \emph{carrier aggregation} (CA) are the two modes of connectivity that can be used to support UE operation over unlicensed spectrum. In the DC mode, a UE can exchange data with multiple gNBs/eNBs simultaneously, where one gNB/eNB is considered the primary and the others as secondary ones. Both primary and secondary gNBs/eNBs connect directly with the core network. 3GPP defines bands of operation in which multiple carriers can be initiated. Under the CA mode, a UE exchanges data with a single gNB/eNB through two or more contiguous or non-contiguous component carriers that could be intra-band or inter-band. For intra-band CA, both primary and secondary carriers are located within the same band, while in the inter-band CA, they can be on different bands. The CA mode enhances the throughput while the DC mode enhances both throughput and reliability, but comes with the complexity of associating a UE with multiple cells. In the DC mode, the failure of the master link does not impact secondary links. Depending on whether DC and/or CA is used to connect with UEs over unlicensed carriers, 3GPP offers flexible NR-U deployment options as explained next (see Figure \ref{fig:deploymentOptions}):

\begin{itemize}
\item Scenario A (NR/NR-U LAA): CA mode consisting of a licensed carrier served by a 5G NR cell and an unlicensed carrier served by a 5G NR-U cell.

\item Scenario B (LTE/NR-U DC): DC mode consisting of a licensed carrier served by an LTE cell and an unlicensed carrier served by a 5G NR-U cell.

\item Scenario C (NR-U Standalone): Standalone mode consisting of unlicensed carrier(s) served by a 5G NR-U cell. This scenario is useful for operating private networks.

\item Scenario D (NR/NR-U UL/DL): Combination of a licensed carrier served by a 5G NR cell for UL communication with an unlicensed carrier served by a 5G NR-U cell for DL communication.

\item Scenario E (NR/NR-U DC): DC mode consisting of a licensed carrier served by a 5G NR cell and an unlicensed carrier served by a 5G NR-U cell.

\end{itemize}

\subsection{Radio Stack} \label{sc:radio-stack}
\begin{figure*}
\centering
\includegraphics[scale=0.65]{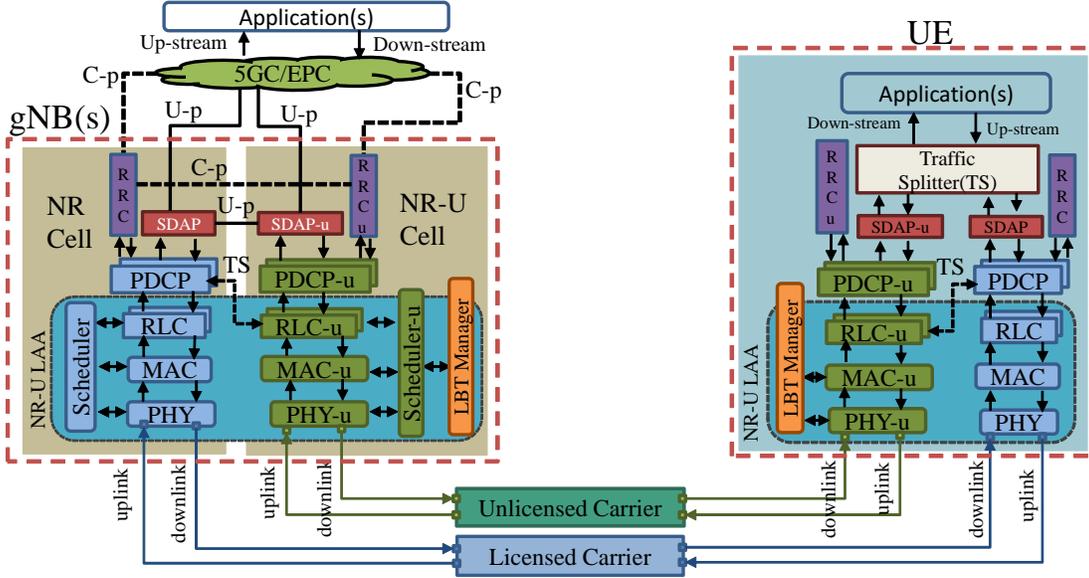}
\caption{NR/NR-U radio stack architecture (Suffix `-u' indicates NR-U block; `U-p': User plane; `C-p': control plane).} \label{fig:radio-stack}
\end{figure*}

To ensure low cost and complexity, as well as easy integration and convergence between NR and NR-U services, the radio stack architecture of NR-U is built upon the NR radio stack, with limited modifications. In this paper, we consider a potential radio stack architecture for a NR/NR-U gNB and a UE, as shown in Figure \ref{fig:radio-stack}. We add suffix `-u' to distinguish NR-U blocks from NR ones. The NR radio stack consists of multiple layers and functional blocks, including the `radio resource control' (RRC), `service data application protocol' (SDAP), `packet data convergence protocol' (PDCP), `radio link control' (RLC), MAC, and PHY. More details on the NR radio stack arhitecture can be found in \cite{TS38.300-nr}. The \emph{LBT Manager} block was added to perform the CATx-LBT procedures, as discussed in Section \ref{sc:lte-felaa-nr-u-overview}. Note that some NR-U blocks may not be present in certain deployment scenarios. For instance, `RRC-unlicensed' (RRC-u), `SDAP-unlicensed' (SDAP-u), `PDCP-unlicensed' (PDCP-u), and `RLC-unlicensed' (RLC-u) are required only for NR-U standalone and NR-U DC-based deployment scenarios. Strategies for traffic splitting/convergence between NR-U and other radio access networks (RANs) can be integrated as part of the `Traffic Splitter' (TS) block. Depending on the NR-U deployment scenario, the TS block can be placed at different levels of the radio stack, as shown in Figure \ref{fig:radio-stack}.

\subsection{Transmission and Signal Design}

\textcolor{black}{In NR-U, the gNB-initiated COT can be split into DL and UL bursts, as shown in Figure \ref{fig:arbitraryTimeOperation}(d). UEs receive and send their control messages within the PDCCH and PUCCH channels. They receive and send their data messages within PDSCH and PUSCH channels. NR-U supports flexible setting of UL and DL allocations in the same COT.} It is a dynamic \emph{time division duplex} (TDD) design in which several DL and UL occasions can take place in a gNB-initiated or a UE-initiated COT. Switching between DL and UL transmissions might be delayed due to the processing required at UE/gNB. If the transition time between DL and UL transmissions, i.e., $T_{\text{sw}}$ in Figure \ref{fig:arbitraryTimeOperation}(d), is longer than $16$ microseconds, UEs should perform CAT2-LBT for $T_{\text{fixed}} = 25$ microseconds before starting their UL transmissions. UEs in a given network can have varying capabilities, and thus proper UL and DL scheduling as well as frame format design are required for efficient COT utilization by considering all UE categories.

 The NR-U uses the same waveform as in NR design. The waveform is an OFDM-modulated signal and has scalable numerology in which multiple subcarrier spacings (SCSs) can be supported, i.e., $\Delta f_\mu  = 15\times 2^{\mu} $ KHz, where $\mu \in\{0,1,2,3\}$ is the numerology index. \textcolor{black}{For operation over UNII bands, 30 KHz is the default SCS.} Every $12$ subcarriers over one OFDM symbol constitute a \emph{resource block} (RB). 
 Multiple numerologies can be multiplexed in the frequency domain using the \emph{bandwidth part} (BWP) concept, in which every BWP has its independent signaling and numerology structure.

  A block of control and data messages, a.k.a., \emph{transport block (TB)}, is transmitted on a time period known as \emph{transmission time interval (TTI)}. Up to two TBs are sent in a TTI. Each TB consists of a set of control messages that are sent over the PDCCH/PUCCH, along with data messages that are sent as part of the PDSCH/PUSCH. Data acknowledgement (ACK) and retransmissions are managed on a TB basis. It is also possible to assign feedback on a codeblock group basis (i.e., segments of TB). Every TB is assigned a unique HARQ process that monitors ACK feedback and handles the retransmission of failed messages. It should be noted that due to the processing delay, it may take the UE/gNB several TTIs before sending the ACK. Therefore, multiple HARQ processes could be active simultaneously and work in parallel to support continuous TB transmissions over time. The granularity of a TTI can be as small as one \emph{mini-slot} or \emph{slot}. Similar to 5G NR, NR-U transmissions are structured into `time slots', each  consisting of $14$ OFDM symbols. The slot duration is $T_{\text{slot}}  = 2^{-\mu}$ milliseconds. In NR-U, it is also possible to have a `mini-slot' that consists of $2$ to $13$ OFDM symbols, which is intended to align NR-U slots with NR slot boundaries.

  \textcolor{black}{It is possible that channel sensing interval may not align with OFDM symbol boundary. In these cases, cyclic prefix (CP) extension can be used to achieve perfect alignment. CP extension can also be used to give UEs more time before switching from DL to UL. CP extension on UL is controlled and configured by RRC layer.}

\section{NR-U Challenges} \label{sc:nru-challenges}
 
\subsection{Interlace Waveform Design}

\begin{figure}
\centering
\includegraphics[scale=0.65]{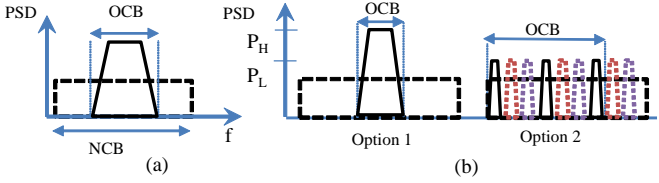}
\caption{(a) Relation between OCB and NCB, (b) examples of NR-U interlace waveform designs (option 1 does not meet the FCC OCB/NCB requirement, but option 2 does)(option 2: Data for each UE is sent every third resource; the solid lines indicate allocations for one UE; the dashed lines indicate potential allocations of another two UEs).} \label{fig:ncb-interlace}
\end{figure}

In UNII bands, the minimum \emph{nominal channel bandwidth (NCB)} is $20$ MHz. The \emph{occupied channel bandwidth (OCB)} , defined as the bandwidth within which $99\%$ of signal power is located. According to the European Telecommunications Standards Institute (ETSI) specifications, OCB should be at least $80\%$ of the NCB. This is needed to achieve harmonious coexistence with other systems, such as Wi-Fi. An example of NCB and OCB is shown in Figure \ref{fig:ncb-interlace}(a). NR-U uses OFDMA for UL transmissions, where different UEs can be scheduled on orthogonal RB resources. Scheduling multiple UEs as well as mixing of different BWPs should be handled carefully to meet the OCB/NCB requirement. For instance, different \emph{interlace waveform} designs, one of which is shown in Figure \ref{fig:ncb-interlace}(b) as Option 2, may be used to satisfy the OCB/NCB requirement. \textcolor{black}{NR-U supports up to 5 and 10 interlace structures for 30 KHz and 15 KHz SCSs, respectively. Additional details on NR-U interlace design can be found in \cite{TS38.211-phy-c-mod}.}


\subsection{Operation Over Multiple Channels}

\textcolor{black}{UNII bands are composed of basic channels of 20 MHz bandwidth. Operating on wider bandwidth can be achieved by bonding channels together.}  \textcolor{black}{NR-U also allows operation on a single wideband carrier that could overlap with a  set of unlicensed channels.} For both standalone and non-standalone modes, NR-U supports operation over a wide bandwidth, composed of a primary channel and multiple secondary channels (a.k.a., multi-carrier channel access) in both standalone and non-standalone modes. \textcolor{black}{ Multi-carrier channel access in LTE-LAA and 5G NR-U is implemented through carrier aggregation.  Compared to LAA, NR-U supports a standalone operation over the unlicensed bands, and thus differs from LAA in that the master and secondary carriers can be both in the unlicensed spectrum. For the non-standalone mode, LTE implements multi-carrier operation over unlicensed bands, a.k.a., \emph{supplemental downlink} and \emph{supplemental uplink}, where a master cell (MCell) operates over a licensed carrier and secondary cells (SCells) are configured to run over unlicensed carriers. NR-U, on the other hand, is supposed to support a standalone operation over unlicensed bands, and thus both MCell and SCell can operate over unlicensed carriers. Options of multi-carrier operation for LAA overlap with the options offered under NR-U. LAA defines two types of channel access over multiple carriers: Type A and Type B. In Type A, a base station conducts individual backoff instance with CAT4-LBT procedure per carrier before accessing it. The base station can access any carrier once the backoff counter of its backoff instance reaches zero. Two subcategories for Type A multi-carrier access are available: Type A1 and Type A2. In Type A1, the backoff counters of different carriers are initiated with different values, while in Type A2, they are all initialized with a common value. In Type B, on the other hand, the base station can access a group of channels simultaneously, where it conducts CAT4-LBT over one carrier and CAT2-LBT over the remaining carriers. The base station simultaneously accesses the cleared carriers. However, the base station is supposed to frequently change the carrier for which it performs CAT4-LBT  (e.g., frequency hopping). Depending on when the base station doubles its contention window, two subcategories of Type B channel access are also defined, namely, Type B1 and Type B2. In Type B1, a common $\text{CW}_{\min}$ value is maintained over all carriers, while in Type B2, different $\text{CW}_{\min}$ values are defined for different carriers. NR-U MIMO operation within NR-U's COT is transparent to the LBT procedure. However, the  energy sensing threshold should be configured properly if the sensed signal is captured over multiple antennas. In other words, the sensing outcomes should not be biased by any applicable analog/digital beamforming}.

\begin{figure*}
\begin{center}
\includegraphics[scale=0.65]{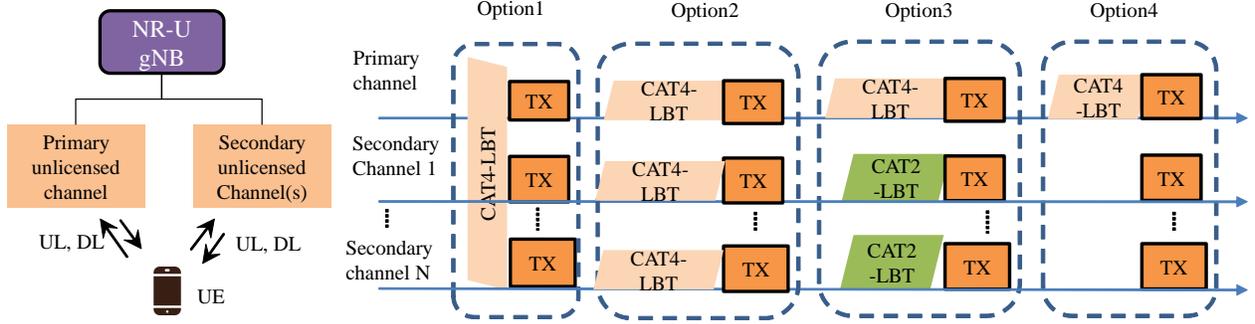}
\caption{Proposed options for standalone NR-U operation on multiple unlicensed channels.} \label{fig:multichannel-lbt}
\end{center}
\end{figure*}

\textcolor{black}{NR-U inherits the multi-carrier access options of LAA.} Multi-carrier channel access is challenging because it is not clear how LBT should be handled when both primary and secondary carriers are configured over unlicensed channels. In Figure \ref{fig:multichannel-lbt}, we propose four options for the LBT procedure in multi-channel access scenario. \textcolor{black}{In Option 1, base station performs a wideband LBT on all channels (i.e., wideband channel) and only starts a COT if all channels are cleared simultaneously. A common counter can be initiated and maintained during the backoff process. Contending on a wide channel bandwidth requires properly setting the sensing threshold to maintain harmonious coexistence with other systems. In Option 2,  gNB/UEs perform CAT4-LBT on all channels separately, and they only access the channels that have been cleared. Option 2 overlaps with Type A of LAA. In Option 3, there could be a single counter maintained for the primary channel, while the secondary channels are sensed for a fixed duration with CAT2-LBT. Option 3 overlaps with Type B of LAA. In Option 4, gNB/UEs perform LBT only on the primary channel and access all channels whenever the primary one is cleared. The forth option is the simplest, but it is expected to create more collisions, especially when these secondary channels are configured as primary for other systems sharing them.} More investigations are required to evaluate the impact of these options on other systems coexisting with NR-U over sophisticated deployments.

\subsection{Frequency Resue for NR-U/Wi-Fi Coexistence}
One of the challenges in NR-U design is how to set the energy detection thresholds, which are used to infer channel occupancy. Energy detection (ED) can provide only a binary indication of the channel status, while preamble detection (PD) provides information on the type of the device occupying the channel. By knowing this type, the detection threshold can be adapted to reduce the impact of hidden/exposed terminals. Adapting the detection threshold can also be done to achieve improved fairness between coexisting technologies. NR-U uses $-72$ dBm as the baseline ED threshold for 20 MHz channel bandwidth. \textcolor{black}{In the absence of other technologies, ED threshold can be relaxed up to -62 dBm. ED threshold needs to be scaled based on the bandwidth of the channel as discussed in \cite{TS37.213-nru}. UEs can be configured by the RRC layer to adapt their ED thresholds while not exceeding regional spectrum regulations.}  Adapting this threshold to improve the spatial frequency reuse and reduce hidden terminals requires more investigation. \textcolor{black}{Reinforcement learning can be a key enabler for such adaptation \cite{hirzallah2020-dissertaion}\cite{Hirzallah2020-icnc}.}  NR-U cells could benefit from Radio Resource Management (RRM) and Radio Link Monitoring (RLM) procedures to better coordinate channel access and assignment among NR-U cells. The same concept can be leveraged by Wi-Fi networks to coordinate their channel access through AP clustering and statistics monitoring. \textcolor{black}{To enable cross-technology signal detection, NR-U and Wi-Fi can benefit from the CP-based signal detection scheme presented in \cite{Hirzallah2017jsac, Hirzallah2016globecom}.}

\subsection{Scheduler and HARQ Design}
\begin{figure}
\centering
\includegraphics[scale=0.5]{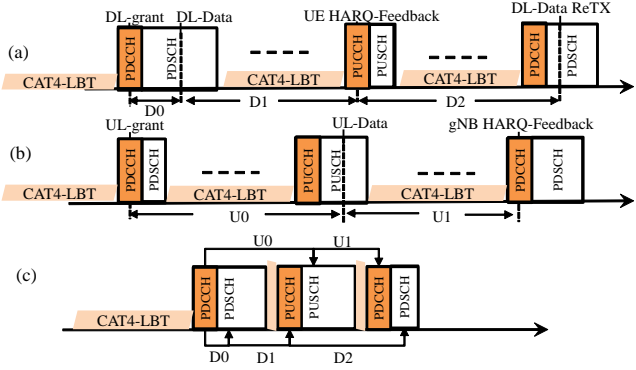}
\caption{NR-U HARQ timing: (a) HARQ timing for DL transmission, (b) HARQ timing for UL transmission, and (c) embedding HARQ control messages within the same COT.} \label{fig:harq-timing}
\end{figure}

DL and UL scheduling in NR-U is asynchronous, meaning that the time to retransmit a failed TB (or codeblock group) is not predetermined and must be indicated explicitly. There are three timing delays governing the dynamics for a DL HARQ process: $D_0$, $D_1$, and $D_2$, as shown in Figure \ref{fig:harq-timing}(a). $D_0$ is the time between a DL grant and DL data occasions. $D_1$ is the time between the DL data and a HARQ feedback message occasions. $D_2$ is the time between HARQ feedback message and data retransmission occasions. In the uplink, there are two key delays, $U_0$ and $U_1$, that govern the timing for the UL HARQ process, as shown in Figure \ref{fig:harq-timing}(b). $U_0$ is the time delay between the notification for an UL grant and the UL data occasions. $U_1$ is the time delay between the UL data and the UL HARQ feedback message occasions. Proper setting of these time delays is critical for harmonious coexistence of NR-U and Wi-Fi systems. Failing to meet these timing constraints can trigger many unnecessary retransmissions on the NR-U side, reducing its throughput and the airtime available for Wi-Fi devices. Configuring these times to be within the gNB-initiated/UE-initiated COT boundary ensures consistent operation for NR-U systems, as shown in Figure \ref{fig:harq-timing}(c). 

The TB can be divided into smaller sub-blocks called the \emph{code block groups} (CBGs), which can be coded individually. In addition to the TB-based HARQ design, NR-U supports a HARQ design for CBGs that takes place at the PHY layer. The impact of TB-based and CBG-based HARQ designs on harmonious NR-U/Wi-Fi coexistence is a topic that requires further investigation.  \textcolor{black}{Enhancements to HARQ include the use of dynamic codebook in which multiple PDSCH occasions (possibly occurring across multiple COTs) can be acknowledged in one codebook feedback message. There is also an option to indicate ACK feedback timing to `later', which means UE can send feedback over coming COTs. Another enhancement is the inclusion of one-shot feedback request whereby gNB can trigger UEs to report all their feedback for all HARQ processes using one-shot feedback report. To distinguish PDSCH occasions that could span over multiple COTs, PDSCH occasions are indexed using the 2 bits `DL assignment index' (DLI) and one bit `PDSCH group'. The gNB indicates its success of receiving  ACK feedback from UE by toggling the New Feedback Indicator (NFI) bit in the DL control signaling. Additional details on signaling of DL and UL resource mapping and allocation can be found in \cite{TS38.213-phy-control}. UL scheduling in NR-U is enhanced to account for unreliable unlicensed channels. A single UL grant can be used to schedule multiple UL TBs for the same UE. Configured UL grant can be sent to allow UEs to autonomously access some UL resources without grant, reducing signaling overhead and providing them more opportunity to access the channel. NR-U also gives more flexibility for sending SRS on any applicable UL OFDM symbol to facilitate wideband channel estimation.}

\subsection{Initial Access and Discovery Design}

Initial access is handled as part of the RRC layer. A UE attaches to the gNB with the highest received power and maintains time/frequency synchronization with it. \textcolor{black}{To send critical messages, such as the \emph{discovery bursts}, the gNB performs CAT2-LBT procedure, as shown in Figure \ref{fig:arbitraryTimeOperation}(e). The DL part of a discovery frame within COT contains the \emph{synchronization signal block} (SSB) burst, which includes initial information (i.e., \emph{master information block} (MIB) and pointers to \emph{remaining system information }(RMSI)) required by UEs to attach to the unlicensed cells over an unlicensed carrier. The SSB consists of Physical Broadcast Channel (PBCH) and synchronization signals, i.e., \emph{primary synchronization signal} (PSS) and \emph{secondary synchronization signal} (SSS)}. To discover a cell, a UE monitors the SSB occasion. The discovery frame is sent using CAT2-LBT channel access to ensure fast delivery, enabling quick initial access and discovery. \textcolor{black}{Discovery frame is sent with periodicity of 20 milliseconds and can take place in 10 or 20 candidate locations within \emph{a discovery burst window} of 5 milliseconds depending on SCS. The COT of a discovery frame can be up to 0.5 or 1 milliseconds depending on SCS.}

 After receiving the discovery frame, the UE starts a \emph{random access channel (RACH)} procedure with the best gNB by engaging in a 2-messages or 4-messages handshake procedure depending on their connectivity status (2-messages handshake procedure can be used for enabling seamless handover between gNBs for connected UEs). The RACH procedure could span more than one COT. To maintain harmonious and fair coexistence of NR-U and IEEE 802.11-based systems, the periodicity of the discovery frame and the RACH procedure should be optimized to support proper initial access without causing impairments to coexisting IEEE 802.11 systems. \textcolor{black}{To account for channel unavailability, NR-U has been supported with additional paging occasions. The design of NR-U includes enhancements to better distinguish between repeated channel access failures and radio link failures.}

\section{Discussion and Simulation Results} \label{sc:evaluation} \label{sc:Performance_Evaluation}

\begin{figure}
\centering
\includegraphics[scale=0.55]{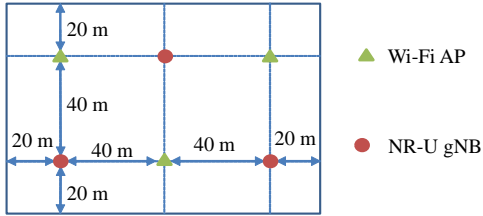}
\caption{3GPP indoor evaluation topology for NR-U/Wi-Fi coexistence \cite{TR38.889-nru}.} \label{fig:topology}
\end{figure}

\begin{figure}
\centering
\includegraphics[scale=0.5]{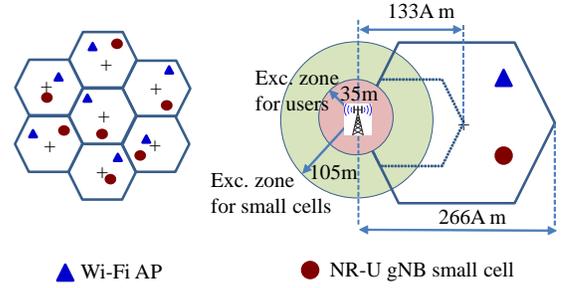}
\caption{3GPP outdoor topology used to evaluate NR-U/Wi-Fi coexistence (A = 1.5) \cite{TR38.889-nru}.} \label{fig:outdoor-topology}
\end{figure}

To evaluate the performance of coexisting NR-U and Wi-Fi networks, we consider an NR-U network that operates according to Scenario D in Figure 3, i.e., a licensed carrier is used for UL communications (via a 5G NR gNB), and an unlicensed carrier is used for DL. We consider an indoor setting in which an NR-U operator deploys three gNBs that share a $20$ MHz channel at $5.18$ GHz with three other IEEE 802.11ac-based APs, as shown in Figure \ref{fig:topology}. Every gNB/AP serves $5$ UEs/STAs, whose locations are randomly selected while ensuring a received power of at least $-82$ dBm. This topology was calibrated and optimized by 3GPP to ensure $10\% $-$ 15 \%$ of received power is below $-72$ dBm, thereby showing the impact of hidden terminals. We consider the the 3GPP InH office pathloss model. We set the transmit power (TP) for gNBs and APs to $23$ dBm, and the TP for UEs and STAs to $18$ dBm. We set the COT/TXOP to $8$ milliseconds, maximum modulation to $64$ QAM, and spatial multiplexiing to 2x2 MIMO for both NR-U and Wi-Fi devices. For the NR-U systems, we set the ED threshold to $-72$ dBm and subcarrier spacing to $15$ KHz. For the Wi-Fi systems, we set the ED threshold to $-62$ dBm and preamble detection threshold to $-82$ dBm. RTS/CTS are disabled. A-MSDU is set to 64 packets. The Minstrel algorithm is used for link adaptation \cite{xia2013minstrel}. \textcolor{black}{We used a customized version of the NS3 simulator (v3.25). We modified the NS3 implementation to accommodate the 3GPP requirements (Study Item TR 38.889 \cite{TR38.889-nru}).} \textcolor{black}{The HARQ design was made more flexible and scalable than in LTE-LAA. HARQ operation and sending of feedback were configured to commence and conclude within one TXOP duration. Traffic generation was brought closer to APs and gNBs. This eliminated the need for backhaul network to connect traffic generators and radio access network, and ensured our results are not biased by delay and scheduling that take place over the backhaul links. To ensure our traffic generation matched real world situation, we implemented traffic generation to be independent and concurrent for both NR-U and Wi-Fi users. Uplink and downlink traffic generations were made independent and concurrent across users.}

We consider FTP traffic that is generated according to the 3GPP FTP model 3 with a file size of $0.5$ MB. Files are generated according to a Poisson process of rate $\lambda$ files per second. NR-U and Wi-Fi devices access the channel using PC $P_3$ and AC\_BE, respectively. For a Wi-Fi user, its traffic is divided equally between UL and DL. For NR-U devices (UEs), the DL traffic is sent over unlicensed spectrum, while the UL traffic is transported over a licensed channel. We only report the DL traffic for NR-U. Each simulation is run for $30$ seconds and repeated $20$ times, where in each time we consider different locations for UEs and STAs. In our simulations, we include the control and management frames used in IEEE 802.11ac and NR-U, and simulate STA association and UE attachment procedures. We study the following performance metrics.

\begin{itemize}
\item \emph{User Perceived Throughput (UPT):} This metric is obtained by dividing the file size in (bits) over its delivery time. Let $t_1$ be the time a file was generated, and let $t_2$ be the time when the last packet from this file was delivered successfully to the receiver. Let $S$ be be the file size in bits. The UPT is computed as $\text{UPT} = S/(t_2 - t_1) $.

\item  \emph{MAC-layer latency} ($T_p$): This is the time needed to deliver a packet between two MAC entities. The latency per packet ($T_p$) includes the queuing time at the transmitter ($T_q$), backoff delay ($T_b$), over-the-air transmission ($T_i$), and processing delays ($T_s$) at the transmitter and receiver. 
 
\item   \emph{Buffer occupancy (BO)}: BO is an indicator of the effectiveness of scheduling and buffer management of various technologies. It is measured by dividing the time for which buffers are non-empty by the total simulation time. 

\item \emph{Utilization factor ($\rho$)}: is obtained by dividing the amount of traffic delivered successfully by the total amount of offered traffic.
\end{itemize}

\subsection{Indoor Coexistence Over the Unlicensed 5 GHz Band}

\begin{figure} 
  \centering
  \includegraphics[scale=.4]{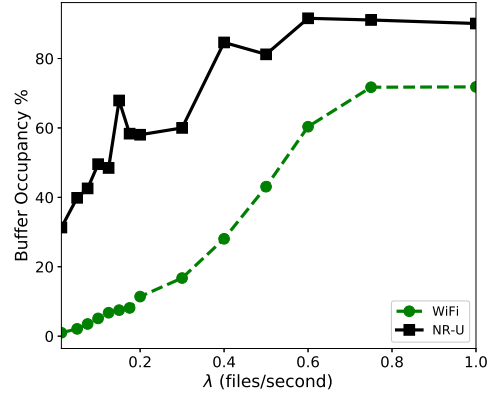}
  \caption{Average buffer occupancy vs. file arrival rate.}  \label{fig:result-bo}
\end{figure}

\begin{figure} 
  \centering
  \includegraphics[scale=.4]{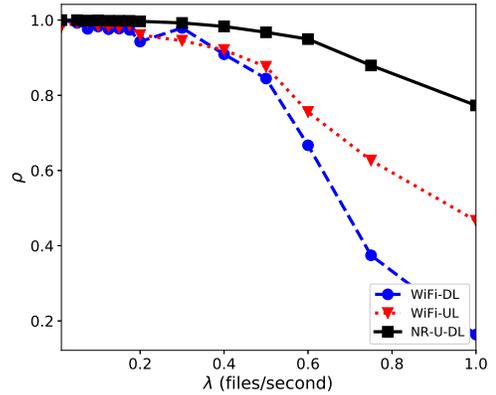}%
  \caption{Average utilization factor vs. file arrival rate.} \label{fig:result-rho}
\end{figure}

 We plot the average BO versus traffic intensity ($\lambda$) in Figure \ref{fig:result-bo}. Under the same $\lambda$, NR-U and Wi-Fi experience different BO behaviors. As $\lambda$ increases, NR-U buffers saturate faster than Wi-Fi. Because of its reliance on OFDMA, NR-U processes buffers and multiplexes UEs differently than Wi-Fi. On the other hand, Wi-Fi usually serves one user at a time. Although NR-U experiences higher BO than Wi-Fi, it is more reliable in terms of packet delivery. We plot $\rho$ versus $\lambda$ in Figure \ref{fig:result-rho}. \textcolor{black}{At low traffic loads, we notice that both NR-U and Wi-Fi have high spectrum utilization. In other words, they both utilize their airtime efficiently and result in many successful transmissions. However, as $\lambda$ increases, the Wi-Fi networks start to experience more losses, dropping many packets due to collisions. On the other hand, NR-U networks seem to be more immune to collisions, providing higher percentage of successful traffic delivery.} At heavy traffic loads, we notice that NR-U has higher spectrum utilization than Wi-Fi and provides more robust and reliable data transmission. \textcolor{black}{NR-U networks perform better in terms of spectrum utilization for several reasons. The design of the HARQ process in NR-U relies on soft-combining, which provides more immunity against interference caused by collisions with Wi-Fi systems. In addition, NR-U takes advantage of the licensed spectrum to exchange critical feedback messages, allowing for timely control of retransmissions. Wi-Fi must rely on unlicensed spectrum to exchange critical messages.}

\begin{figure} 
  \centering
  \includegraphics[scale=.4]{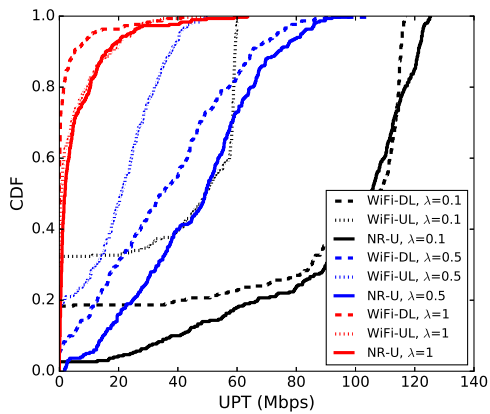}
   \caption{CDF of UPT (indoor at 5.18 GHz; transmit power: 23 dBm; channel bandwidth: 20 MHz; 3GPP InH path loss model).} 
\label{fig:result-upt-cdf}%
  \end{figure}

\begin{figure} 
  \centering
  \includegraphics[scale=.4]{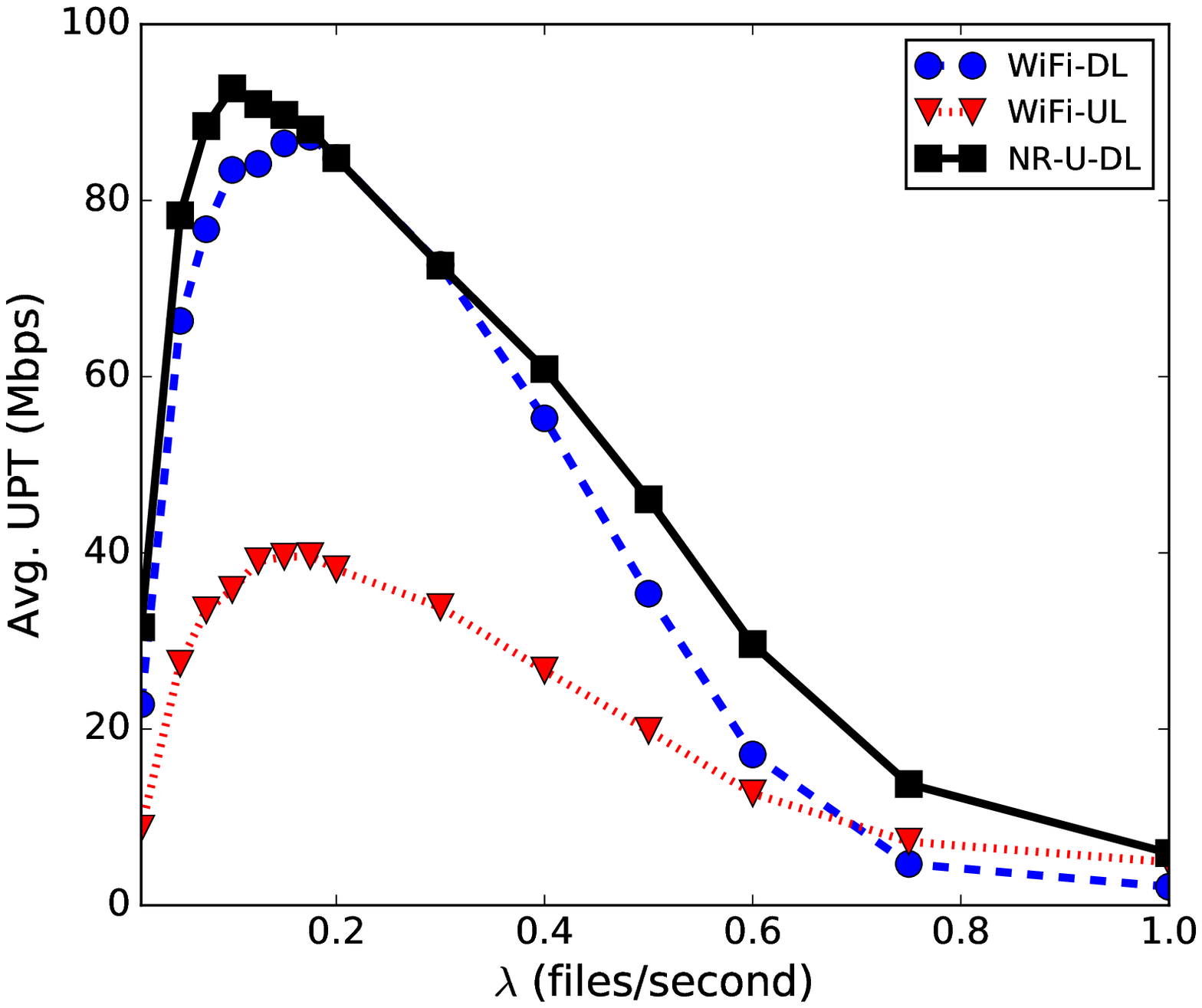}%
  \caption{Average UPT vs. $\lambda$ (indoor at 5.18 GHz; transmit power: 23 dBm; channel bandwidth: 20 MHz; 3GPP InH path loss model).} \label{fig:result-upt}%
  \end{figure}

In Figure \ref{fig:result-upt-cdf}, we plot the CDF of the UPT for Wi-Fi DL and UL, as well as NR-U DL. We also plot the average UPT versus $\lambda$ in Figure \ref{fig:result-upt}. NR-U maintains higher average UPT than Wi-Fi UL and DL. At heavy traffic loads, we notice that Wi-Fi STAs experience outage, where about 30\% of users receive zero throughput. This happens because some critical Wi-Fi control and management frames are lost due to collisions with NR-U transmissions. For instance, we noticed many failed attempts to deliver reassociation frames and/or frames carrying important messages, such as address resolution protocol (ARP) request/reply messages. These issues did not happen in NR-U because NR-U uplink traffic goes overa licensed channel. \textcolor{black}{Similar observations of Wi-Fi losing frames that contain  important control messages were also reported in other studies (see \cite{Sathya2020-access-lte-u} and  references therein).}

\begin{figure} 
  \centering
  \includegraphics[scale=.4]{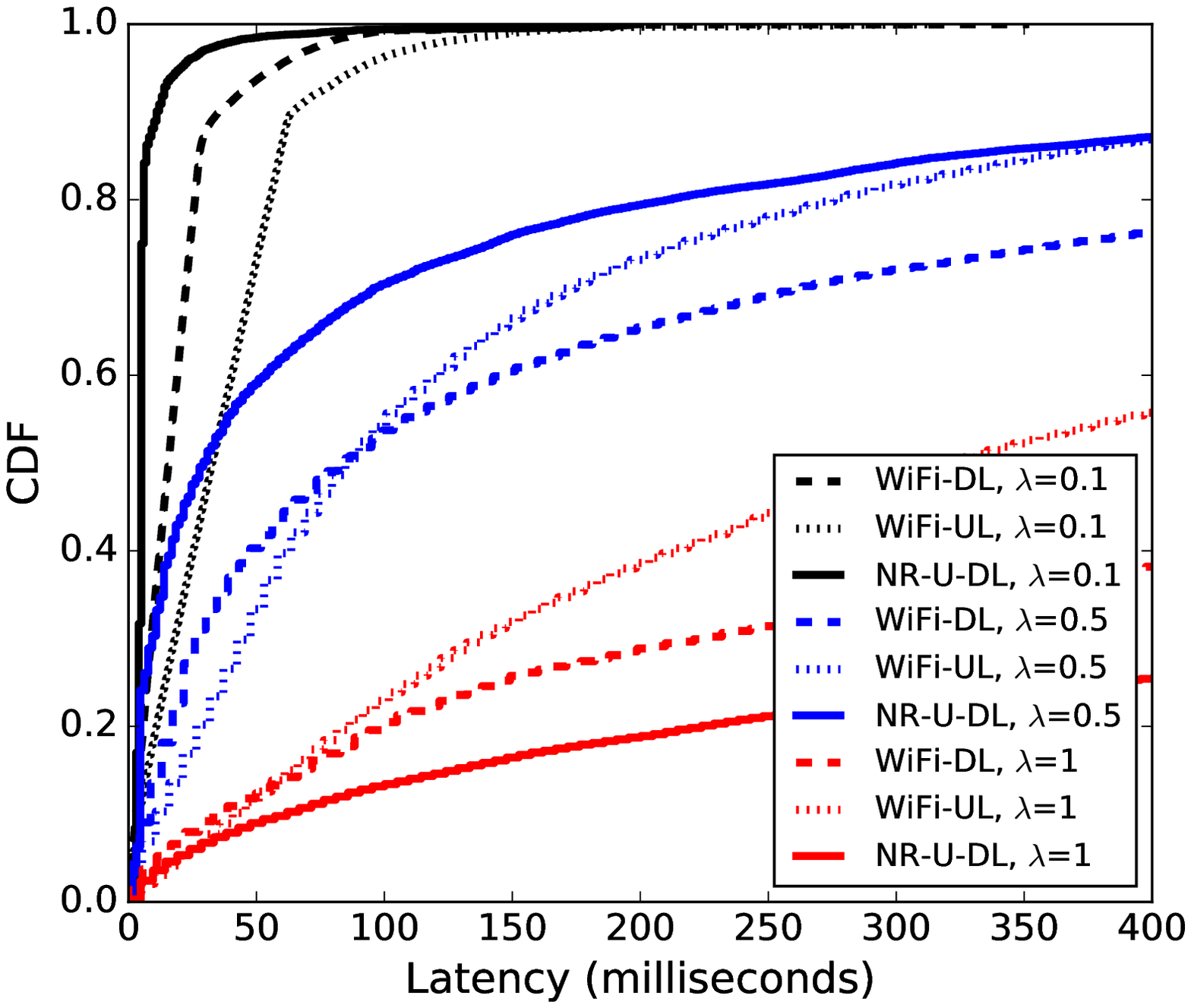}
  \caption{CDF of latency (indoor at 5.18 GHz; transmit power: 23 dBm; channel bandwidth: 20 MHz; 3GPP InH path loss model).} 
  \label{fig:result-latency-cdf}%
  \end{figure}

\begin{figure} 
  \centering
  \includegraphics[scale=.4]{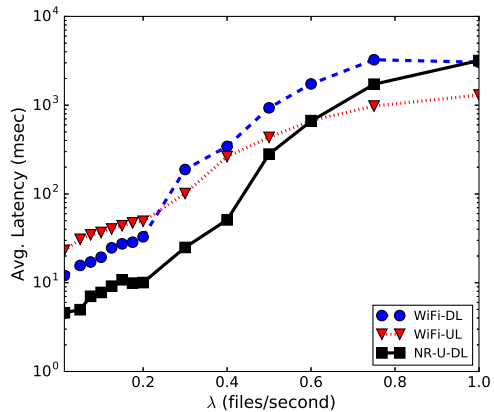}
  \caption{Average latency vs. $\lambda$ (indoor at 5.18 GHz; transmit power: 23 dBm; channel bandwidth: 20 MHz; 3GPP InH path loss model).} 
  \label{fig:result-latency}
  \end{figure}

We plot the latency CDF in Figure \ref{fig:result-latency-cdf}, and the average latency versus traffic intensity in Figure \ref{fig:result-latency}. \textcolor{black}{The latency measurements are reported on a per packet basis. It can be observed that Wi-Fi devices experience higher latency than NR-U. At light traffic, both networks experience latency  below 100 milliseconds. However, as $\lambda$ increases, the per-packet latency increases exponentially, with the average latency per packet becoming in the orders of seconds. At heavy load, the average latencies for Wi-Fi and NR-U networks are comparable. As observed from NR-U latency performance, the effectiveness of using NR-U for supporting URLLC applications over unlicensed bands can be challenging. Enabling the use of higher numerology indices for NR-U operation over unlicensed bands could result in lower latency at heavy loads.}

\subsection{Outdoor Coexistence Over the Unlicensed 5 GHz Band}
\begin{figure}
\centering
\includegraphics[scale=0.4]{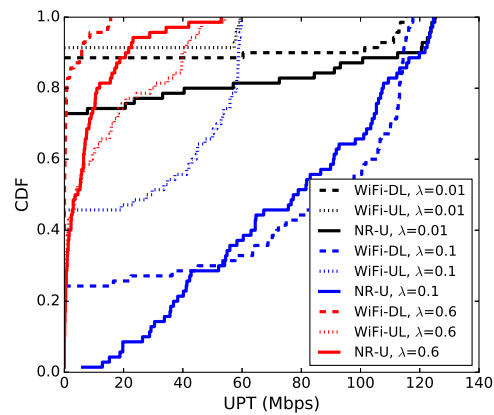}
\caption{CDF of UPT (outdoor at 5.18 GHz; transmit power: 23 dBm; channel bandwidth: 20 MHz; 3GPP UMi street canyon path loss model).} \label{fig:upt-cdf-outdoor}
\end{figure}

\begin{figure}
\centering
\includegraphics[scale=0.4]{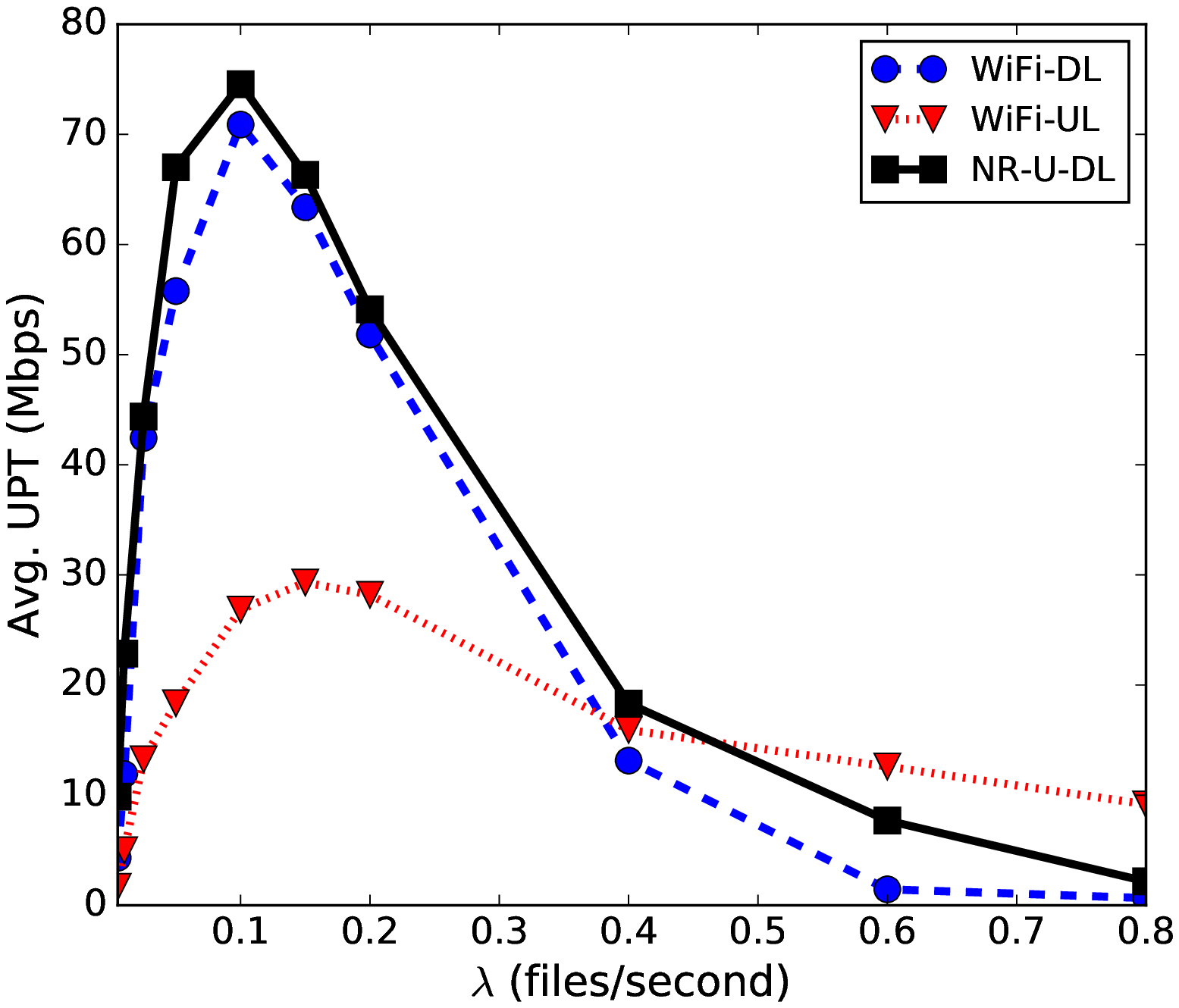}
\caption{Average UPT vs. $\lambda$ (outdoor at 5.18 GHz; transmit power: 23 dBm; channel bandwidth: 20 MHz; 3GPP UMi street canyon path loss model).} \label{fig:upt-lambda-outdoor}
\end{figure}

\begin{figure}
\centering
\includegraphics[scale=0.4]{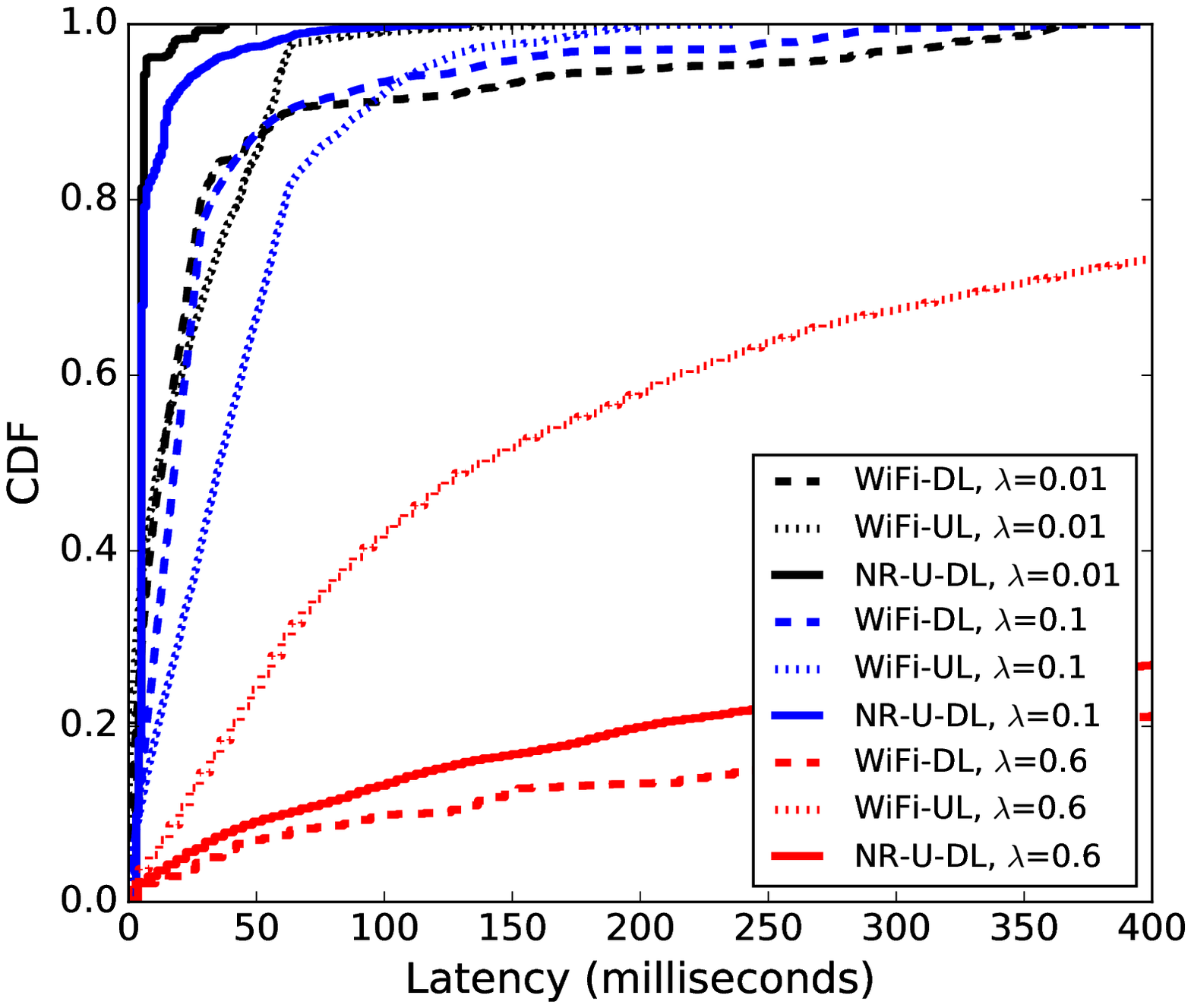}
\caption{CDF of latency (outdoor at 5.18 GHz; transmit power: 23 dBm; channel bandwidth: 20 MHz; 3GPP UMi street canyon path loss model).} \label{fig:latency-cdf-outdoor}
\end{figure}

\begin{figure}
\centering
\includegraphics[scale=0.4]{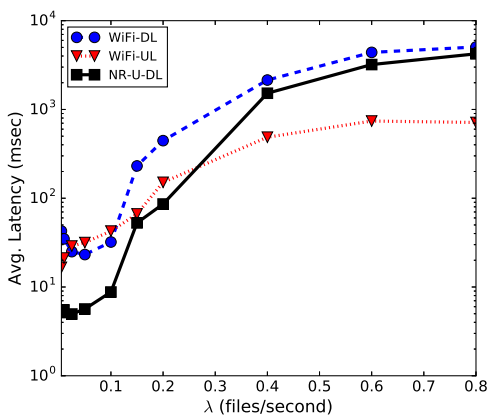}
\caption{Average latency vs. $\lambda$ (outdoor at 5.18 GHz; transmit power: 23 dBm; channel bandwidth: 20 MHz; 3GPP UMi street canyon path loss model).} \label{fig:latency-lambda-outdoor}
\end{figure}

\textcolor{black}{Next, we investigate outdoor scenarios, as shown in Figures \ref{fig:upt-cdf-outdoor}, \ref{fig:upt-lambda-outdoor}, \ref{fig:latency-cdf-outdoor}, and \ref{fig:latency-lambda-outdoor}. We consider 3GPP UMi Street Canyon path loss model and the outdoor topology in Figure \ref{fig:outdoor-topology}. Under light and medium traffic loads, NR-U achieves higher MAC throughput than Wi-Fi. However, under heavy traffic, we observe that NR-U performance becomes worse than Wi-Fi uplink. Both NR-U and Wi-Fi downlink have comparable performance. Under heavy traffic load, Wi-Fi stations become more active and aggressive in accessing the unlicensed channel because of their reliance on higher energy sensing thresholds than NR-U, i.e., $-62$ dBm vs. $-72$ dBm. This results in Wi-Fi uplink achieving higher throughput. The same observation applies to outdoor latency performance, as can be observed in Figures \ref{fig:latency-cdf-outdoor} and \ref{fig:latency-lambda-outdoor}. The latency of NR-U and Wi-Fi are comparable under low and medium traffic loads. Under heavy load, NR-U and Wi-Fi downlink streams experience higher latency than Wi-Fi uplink streams.}

\subsection{Indoor Coexistence Over the Unlicensed 6 GHz bands}
\begin{figure}
\begin{center}
\includegraphics[scale=0.4]{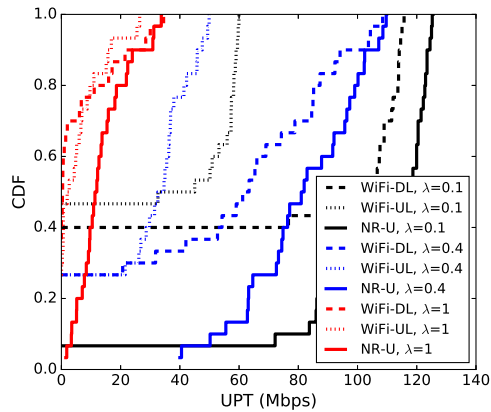}
\caption{CDF of UPT (indoor at 6.18 GHz; transmit power: 18 dBm; channel bandwidth: 20 MHz; 3GPP InH path loss model).} \label{fig:upt-cdf-indoor-6ghz}
\end{center}
\end{figure}

\begin{figure}
\begin{center}
\includegraphics[scale=0.4]{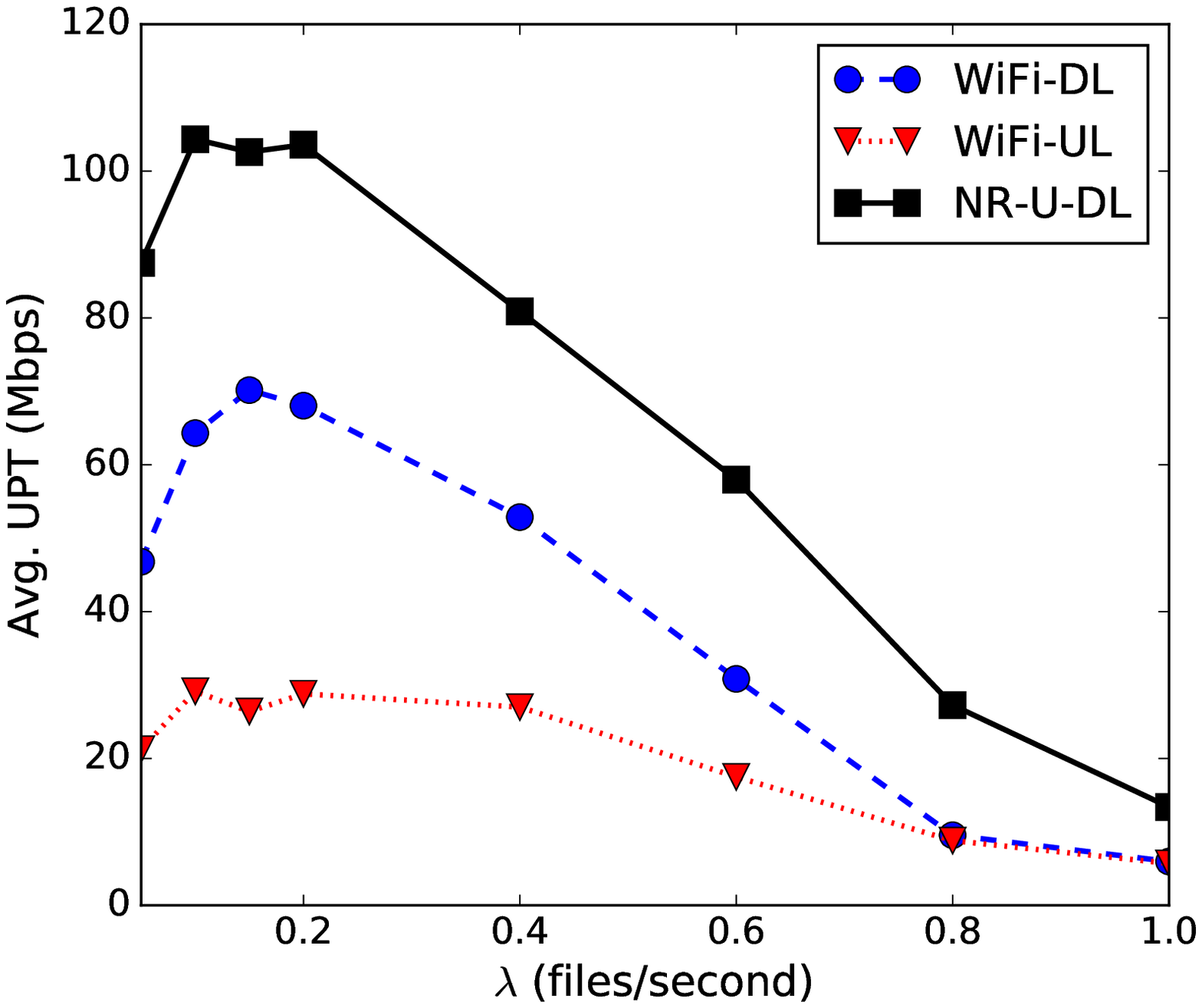}
\caption{Average UPT vs. $\lambda$ (indoor at 6.18 GHz; transmit power: 18 dBm; channel bandwidth: 20 MHz; 3GPP InH path loss model).} \label{fig:upt-lambda-indoor-6ghz}
\end{center}
\end{figure}

\begin{figure}
\begin{center}

\includegraphics[scale=0.4]{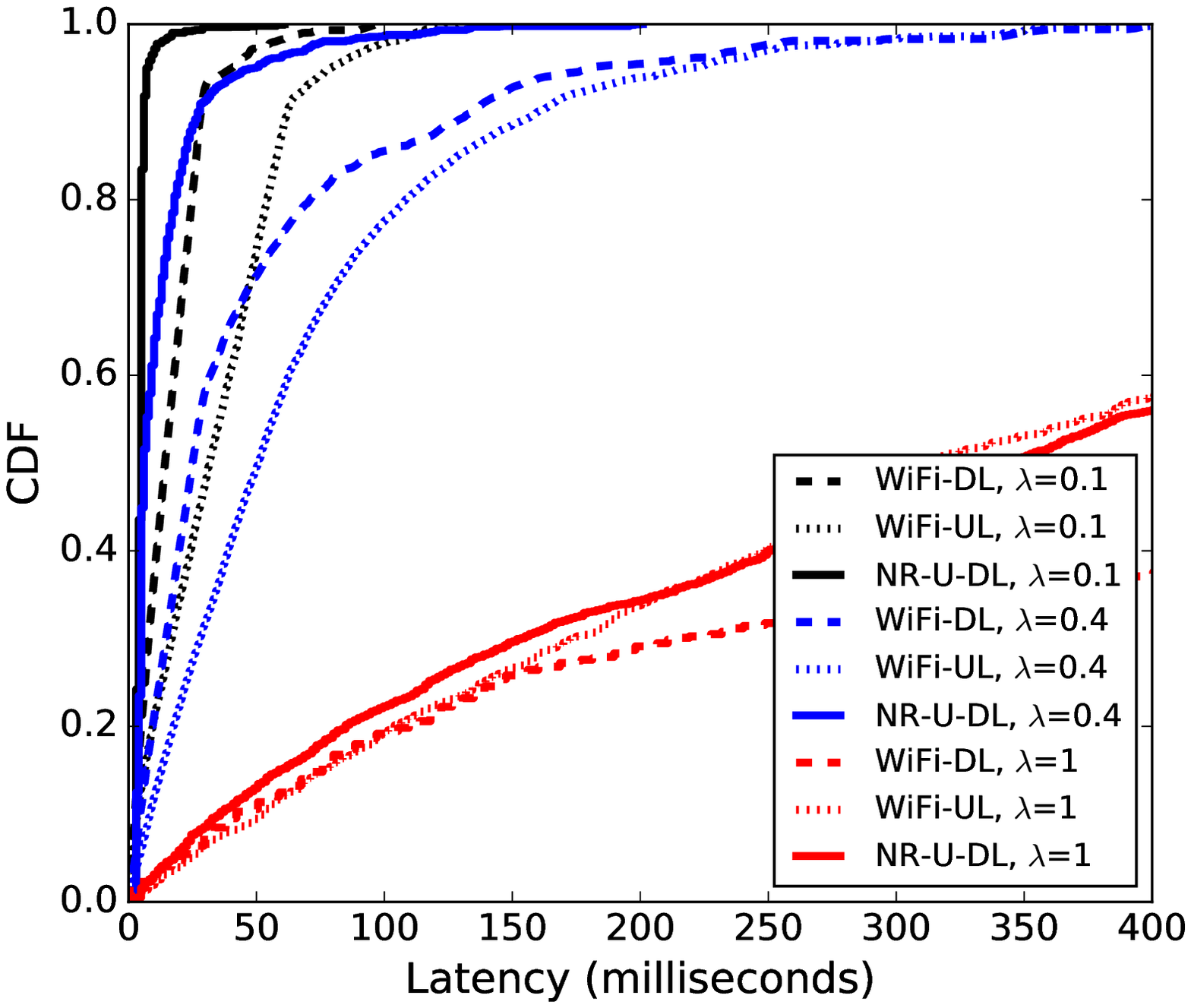}
\caption{CDF of latency (indoor at 6.18 GHz; transmit power: 18 dBm; channel bandwidth: 20 MHz; 3GPP InH path loss model).} \label{fig:latency-cdf-indoor-6ghz}
\end{center}
\end{figure}

\begin{figure}
\begin{center}
\includegraphics[scale=0.4]{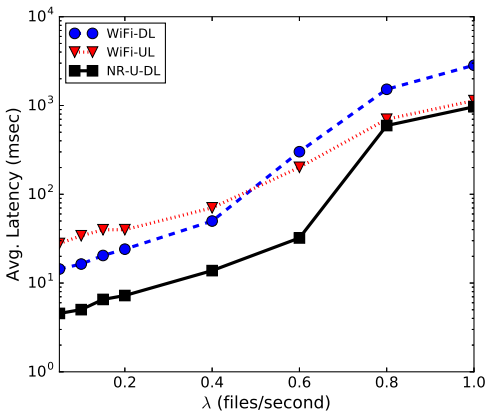}
\caption{Average latency vs. $\lambda$ (indoor at 6.18 GHz; transmit power: 18 dBm; channel bandwidth: 20 MHz; 3GPP InH path loss model).} \label{fig:latency-lambda-indoor-6ghz}
\end{center}
\end{figure}

\textcolor{black}{Following the FCC's most recent Report and Order and FNPRM for operation over the unlicensed 6 GHz bands \cite{FCC_2020-6ghz}, we set the transmit power of base stations to $18$ dBm and that of users to $12$ dBm (see Table 3 \cite{FCC_2020-6ghz}). We report the per-user MAC throughput and per-packet latency for indoor operation at $6.18$ GHz center frequency. The CDF of the per-user MAC throughput is shown in Figure \ref{fig:upt-cdf-indoor-6ghz}, while the average MAC throughput versus traffic rate ($\lambda$) is shown in Figure \ref{fig:upt-lambda-indoor-6ghz}. The rest of simulation parameters are set as in the case of indoor operation over the unlicensed 5 GHz bands. Inline with our previous observations of the performance over the 5 GHz bands, NR-U achieves higher throughput than Wi-Fi. We observe that NR-U average gain over Wi-Fi is even higher than in the case of the unlicensed 5 GHz band. The ratio of NR-U to Wi-Fi highest average throughout is 1.5 at the unlicensed 6 GHz band, while it is 1.17 for the case of unlicensed 5 GHz band. Unlike Wi-Fi, NR-U is less affected by the reduction in the transmit power, and this is due to the fact that NR-U has more sophisticated interference mitigation, rate control, and HARQ designs than Wi-Fi. We also report the CDF for the packet latency in Figure \ref{fig:latency-cdf-indoor-6ghz}, and the average packet latency versus $\lambda$ in Figure \ref{fig:latency-lambda-indoor-6ghz}. As can be observed, NR-U achieves lower latency than Wi-Fi, but the gap in latency shrinks as $\lambda$ increases. In the DL, NR-U achieves a comparable latency performance to Wi-Fi at heavy traffic load, as shown in Figure \ref{fig:latency-lambda-indoor-6ghz}. }

\section{Conclusions} \label{sc:conclusions}
This paper provided an overview of 5G NR-U technology and discussed open challenges to operate it in the presence of Wi-Fi systems. Achieving harmonious NR-U/Wi-Fi coexistence requires investigating many NR-U issues, including waveform design, multi-channel operation, frequency reuse, scheduling and HARQ, as well as the initial access and discovery design. Our simulations indicate that under heavy traffic, NR-U achieves higher throughput and lower latency than Wi-Fi, and experiences different BO and spectrum utilization statistics than Wi-Fi. We found that the loss of certain critical control messages due to collisions with NR-U transmission messages could be detrimental to Wi-Fi operation. \textcolor{black}{ There is a need for additional enhancements to ensure successful delivery of critical Wi-Fi messages. For example, Wi-Fi could be configured with more reliable and fast LBT parameters when sending critical messages. We also found that NR-U effectiveness to support URLLC applications over unlicensed bands is still questionable. NR-U design needs to support additional enhancement for securing lower latency. Examples of these enhancement could be using higher numerology indices and supporting faster retransmissions.}




\bibliographystyle{IEEEtran}
\bibliography{tccn2020}

\begin{thebibliography}{10}
\providecommand{\url}[1]{#1}
\csname url@samestyle\endcsname
\providecommand{\newblock}{\relax}
\providecommand{\bibinfo}[2]{#2}
\providecommand{\BIBentrySTDinterwordspacing}{\spaceskip=0pt\relax}
\providecommand{\BIBentryALTinterwordstretchfactor}{4}
\providecommand{\BIBentryALTinterwordspacing}{\spaceskip=\fontdimen2\font plus
\BIBentryALTinterwordstretchfactor\fontdimen3\font minus
  \fontdimen4\font\relax}
\providecommand{\BIBforeignlanguage}[2]{{%
\expandafter\ifx\csname l@#1\endcsname\relax
\typeout{** WARNING: IEEEtran.bst: No hyphenation pattern has been}%
\typeout{** loaded for the language `#1'. Using the pattern for}%
\typeout{** the default language instead.}%
\else
\language=\csname l@#1\endcsname
\fi
#2}}
\providecommand{\BIBdecl}{\relax}
\BIBdecl

\bibitem{NR-U-WID}
3GPP, ``{NR-based Access to Unlicensed Spectrum},'' no. 3GPP RP-190706, 2019.

\bibitem{Semaan2017wcnc-nru-beamforming}
E.~{Semaan}, J.~{Ansari}, G.~{Li}, E.~{Tejedor}, and H.~{Wiemann}, ``An outlook
  on the unlicensed operation aspects of {NR},'' in \emph{Proc. of IEEE
  Wireless Communications and Networking Conference (WCNC'17)}, March 2017, pp.
  1--6.

\bibitem{Pater2018mag-5g-laa-slicing-huwaei}
E.~{Pateromichelakis}, O.~{Bulakci}, C.~{Peng}, J.~{Zhang}, and Y.~{Xia},
  ``{LAA} as a key enabler in slice-aware {5G RAN}: Challenges and
  opportunities,'' \emph{IEEE Communications Standards Magazine}, vol.~2,
  no.~1, pp. 29--35, March 2018.

\bibitem{Lagen2020nru-mmwave}
S.~{Lagen}, L.~{Giupponi}, S.~{Goyal}, N.~{Patriciello}, B.~{Bojović},
  A.~{Demir}, and M.~{Beluri}, ``New radio beam-based access to unlicensed
  spectrum: Design challenges and solutions,'' \emph{IEEE Communications
  Surveys Tutorials}, vol.~22, no.~1, pp. 8--37, 2020.

\bibitem{Oh2019globecom-nru-samsung}
J.~{Oh}, Y.~{Kim}, Y.~{Li}, J.~{Bang}, and J.~{Lee}, ``Expanding {5G} new radio
  technology to unlicensed spectrum,'' in \emph{Proc. of IEEE Globecom'19
  Conf.}, 2019, pp. 1--6.

\bibitem{Lu2019nru-Iot}
X.~{Lu}, V.~{Petrov}, D.~{Moltchanov}, S.~{Andreev}, T.~{Mahmoodi}, and
  M.~{Dohler}, ``{5G-U}: Conceptualizing integrated utilization of licensed and
  unlicensed spectrum for future {IoT},'' \emph{IEEE Communications Magazine},
  vol.~57, no.~7, pp. 92--98, 2019.

\bibitem{TR38.889-nru}
3GPP, ``{Study on NR--based access to unlicensed spectrum},'' no. 3GPP TR
  38.889 v16.0.0, Dec. 2018.

\bibitem{zhao2019nru-phy}
Y.~Zhao, J.~Liu, and S.~Xie, ``Cubic metric reduction for repetitive cazac
  sequences in frequency domain in {5G} system,'' \emph{arXiv preprint
  arXiv:1910.11184}, 2019.

\bibitem{Maldona2020-access}
R.~{Maldonado}, C.~{Rosa}, and K.~I. {Pedersen}, ``Latency and reliability
  analysis of cellular networks in unlicensed spectrum,'' \emph{IEEE Access},
  vol.~8, pp. 49\,412--49\,423, 2020.

\bibitem{Patriciello2020-access}
N.~{Patriciello}, S.~{Lagén}, B.~{Bojović}, and L.~{Giupponi}, ``{NR-U} and
  {IEEE} 802.11 technologies coexistence in unlicensed mmwave spectrum: Models
  and evaluation,'' \emph{IEEE Access}, vol.~8, pp. 71\,254--71\,271, 2020.

\bibitem{hirzallah2020-dissertaion}
\BIBentryALTinterwordspacing
M.~Hirzallah, ``Protocols and algorithms for harmonious coexistence over
  unlicensed bands in next-generation wireless networks,'' Ph.D. dissertation,
  Univ. of Arizona, Tucson, Aug. 2020. [Online]. Available:
  \url{https://repository.arizona.edu/handle/10150/645808}
\BIBentrySTDinterwordspacing

\bibitem{Hirzallah2019Secon-matchmaker}
M.~{Hirzallah}, Y.~{Xiao}, and M.~{Krunz}, ``Matchmaker: An inter-operator
  network sharing framework in unlicensed bands,'' in \emph{to appear in proc.
  of IEEE SECON'19 Conference}, June 2019, pp. 1--9.

\bibitem{kosek2020downlink}
K.~Kosek-Szott, A.~L. Valvo, S.~Szott, I.~Tinnirello, and P.~Gallo, ``Downlink
  channel access performance of nr-u: Impact of scheduling flexibility on
  coexistence with wi-fi in the 5 ghz band,'' \emph{arXiv preprint
  arXiv:2007.14247}, 2020.

\bibitem{FCC-6GHz-18-295}
{Federal Communications Commission}, ``{Fact Sheet: Unlicensed Use of the 6 GHz
  Band Notice of Proposed Rulemaking},'' no. 18-295, Oct. 2018.

\bibitem{FCC-60GHz-340310A1}
------, ``Fact sheet: Spectrum frontiers rules identify, open up vast amounts
  of new high-band spectrum for next generation {(5G)} wireless broadband,''
  no. 340310A1, 2016.

\bibitem{FCC_2020-6ghz}
------, ``{Report and Order and Further Notice of Proposed Rulemaking -
  Unlicensed Use of the 6 GHz Band},'' \emph{FCC}, no. 20-51, 2020.

\bibitem{TR37.890-6ghz}
3GPP, ``{Feasibility Study on 6 {GHz} for {LTE} and {NR} in Licensed and
  Unlicensed Operations (Release 17) },'' no. 3GPP TR 37.890 v0.9.0, Sep. 2020.

\bibitem{naik2020next}
G.~Naik, J.-M. Park, J.~Ashdown, and W.~Lehr, ``Next generation {Wi-Fi} and
  {5G} {NR-U} in the 6 {GHz} bands: Opportunities and challenges,'' \emph{IEEE
  Access}, vol.~8, pp. 153\,027--153\,056, 2020.

\bibitem{IEEE802.11-2016}
{IEEE}, ``{IEEE}--part 11: Wireless lan mac and phy layer specifications,''
  2016.

\bibitem{TR36.213LAA}
3GPP, ``Physical layer procedures,'' 3{GPP} {TR}. 36.213 v15.1.0., Mar. 2018.

\bibitem{TS37.213-nru}
------, ``{Physical layer procedures for shared spectrum channel access},'' no.
  3GPP TS 37.213 v16.3.0, Sep. 2020.

\bibitem{Khorov2019tutorial-11ax}
E.~{Khorov}, A.~{Kiryanov}, A.~{Lyakhov}, and G.~{Bianchi}, ``A tutorial on
  {IEEE} 802.11ax high efficiency {WLANs},'' \emph{IEEE Communications Surveys
  Tutorials}, vol.~21, no.~1, pp. 197--216, 2019.

\bibitem{Hirzallah2018-tccn-fd-wifi}
M.~{Hirzallah}, W.~{Afifi}, and M.~{Krunz}, ``Provisioning qos in {Wi-Fi}
  systems with asymmetric full-duplex communications,'' \emph{IEEE Transactions
  on Cognitive Communications and Networking}, vol.~4, no.~4, pp. 942--953,
  2018.

\bibitem{Lopez2019comm-mag-11be}
D.~{Lopez-Perez}, A.~{Garcia-Rodriguez}, L.~{Galati-Giordano}, M.~{Kasslin},
  and K.~{Doppler}, ``{IEEE} 802.11be extremely high throughput: The next
  generation of {Wi-Fi} technology beyond 802.11ax,'' \emph{IEEE Communications
  Magazine}, vol.~57, no.~9, pp. 113--119, 2019.

\bibitem{Deng2020-11be-tutorial}
C.~{Deng}, X.~{Fang}, X.~{Han}, X.~{Wang}, L.~{Yan}, R.~{He}, Y.~{Long}, and
  Y.~{Guo}, ``{IEEE} 802.11be {Wi-Fi} 7: {N}ew challenges and opportunities,''
  \emph{IEEE Communications Surveys Tutorials}, vol.~22, no.~4, pp. 2136--2166,
  2020.

\bibitem{Khorov2020-access-11be}
E.~{Khorov}, I.~{Levitsky}, and I.~F. {Akyildiz}, ``Current status and
  directions of {IEEE} 802.11be, the future {Wi-Fi} 7,'' \emph{IEEE Access},
  vol.~8, pp. 88\,664--88\,688, 2020.

\bibitem{Hirzallah2019tccn-modeling}
M.~{Hirzallah}, M.~{Krunz}, and Y.~{Xiao}, ``Harmonious cross-technology
  coexistence with heterogeneous traffic in unlicensed bands: Analysis and
  approximations,'' \emph{IEEE Transactions on Cognitive Communications and
  Networking}, vol.~5, no.~3, pp. 690--701, 2019.

\bibitem{Hirzallah2018Dyspan-laa-modeling}
M.~{Hirzallah}, Y.~{Xiao}, and M.~{Krunz}, ``On modeling and optimizing
  lte/wi-fi coexistence with prioritized traffic classes,'' in \emph{Proc. of
  IEEE International Symposium on Dynamic Spectrum Access Networks (DySPAN)},
  Oct 2018, pp. 1--10.

\bibitem{sathya2020measurement}
V.~Sathya, M.~I. Rochman, and M.~Ghosh, ``Measurement-based coexistence studies
  of {LAA} \& {Wi-Fi} deployments in {C}hicago,'' \emph{arXiv preprint
  arXiv:2010.15012}, 2020.

\bibitem{TS38.300-nr}
3GPP, ``{NR and NG--RAN Overall Description --Stage 2},'' no. 3GPP TS 38.300
  v15.4.0, Dec. 2018.

\bibitem{TS38.211-phy-c-mod}
------, ``{NR} physical channels and modulation,'' no. 3GPP TS 38.211 v16.3.0,
  Sep. 2020.

\bibitem{Hirzallah2020-icnc}
M.~{Hirzallah} and M.~{Krunz}, ``Intelligent tracking of network dynamics for
  cross-technology coexistence over unlicensed bands,'' in \emph{Proc. of the
  IEEE International Conference on Computing, Networking and Communications
  (ICNC)}, 2020, pp. 698--703.

\bibitem{Hirzallah2017jsac}
M.~Hirzallah, W.~Afifi, and M.~Krunz, ``Full-duplex-based rate/mode adaptation
  strategies for {Wi-Fi/LTE-U} coexistence: A {POMDP} approach,'' \emph{IEEE
  Journal on Selected Areas in Communications}, vol.~35, no.~1, pp. 20--29, Jan
  2017.

\bibitem{Hirzallah2016globecom}
------, ``Full-duplex spectrum sensing and fairness mechanisms for
  {Wi-Fi/LTE-U} coexistence,'' in \emph{Proc. of the {IEEE GLOBECOM'16} Conf.},
  Dec 2016, pp. 1--6.

\bibitem{TS38.213-phy-control}
3GPP, ``{NR} physical layer procedures for control,'' no. 3GPP TS 38.213
  v16.3.0, Sep. 2020.

\bibitem{xia2013minstrel}
D.~Xia, J.~Hart, and Q.~Fu, ``Evaluation of the minstrel rate adaptation
  algorithm in ieee 802.11 g wlans,'' in \emph{Proc. of IEEE International
  Conference on Communications (ICC)}.\hskip 1em plus 0.5em minus 0.4em\relax
  IEEE, 2013, pp. 2223--2228.

\bibitem{Sathya2020-access-lte-u}
V.~{Sathya}, M.~{Mehrnoush}, M.~{Ghosh}, and S.~{Roy}, ``{Wi-Fi/LTE-U}
  coexistence: Real-time issues and solutions,'' \emph{IEEE Access}, vol.~8,
  pp. 9221--9234, 2020.

\end{thebibliography}
%




\end{document}